\documentclass{nature}
\usepackage{epsfig}
\usepackage{amsmath,mathrsfs,amsbsy,color,graphicx,bm,amsthm,amsfonts}

\begin{document}

\title{Quantum-enhanced metrology for multiple phase estimation with noise}
\author{Jie-Dong Yue$^1$, Yu-Ran Zhang$^1$, and Heng Fan$^{1,2,\star}$}
\maketitle

\begin{affiliations}
\item
Beijing National Laboratory for Condensed Matter Physics, Institute of
Physics, Chinese Academy of Sciences, Beijing 100190, China

\item
Collaborative Innovation Center of Quantum Matter, Beijing 100190, China

$^\star$e-mail: hfan@iphy.ac.cn;

\end{affiliations}

\begin{abstract}
We present a general quantum metrology framework to study the simultaneous estimation of multiple phases
in the presence of noise as a discretized model for phase imaging.
This approach can lead to nontrivial bounds of the precision for multiphase estimation.
Our results show that simultaneous estimation (SE) of multiple phases is always better than individual estimation (IE) of each phase
even in noisy environment.
The utility of the bounds of multiple phase estimation for photon loss channels is exemplified explicitly.
When noise is low, those bounds possess the Heisenberg scale showing
quantum-enhanced precision with the $O(d)$ advantage for SE,
where $d$ is the number of phases.
However, this $O(d)$ advantage of SE scheme in the variance of the estimation may disappear asymptotically when
photon loss becomes significant and then only a constant advantage over that of IE scheme demonstrates.
Potential application of those results is presented.
\end{abstract}

A general estimation scheme of multiple parameters can be divided into three stages: the preparation of some probes, the interaction of the probes with a system which is determined by the parameter vector $\bm{\theta}$, and measurements of the probes after the interaction. Then $\bm{\theta}$ is estimated from the results of the measurements. When the dimension of $\bm{\theta}$ is 1, the case becomes single parameter estimation. If the probes are uncorrelated, then the central limit theorem states that the estimation error $Tr[Cov(\bm{\theta})]$ scales as $1/\sqrt{N}$, with $N$ being the number of resources (photons, atoms) employed. While in quantum world by correlating the probes nonclassically,
the estimation error may scale as $1/N$ in an ideal scenario, which is the ultimate limit of precision named as the Heisenberg limit \cite{wineland1992spin, Wineland1994, giovannetti2006quantum, giovannetti2004quantum}.
The enhancement in the estimation precision is the main concern
of quantum metrology, and a lot of work has been done, both theoretically and experimentally \cite{Huelgal1997,boixo2007generalized,monras2007optimal,dorner2009optimal,higgins2007entanglement,konrad2009quantum,
m2010experimental,caves1981quantum,andre2004stability,berry2000optimal,armen2002adaptive,hentschel2010machine,LloydReview,
XiangNatPhoton,ZhangYLPRA,ZhangYLPRA2,Lukin1,Lukin2,PhysRevLett.109.123601,Humphreys2013,gill2000state,cyril2013}

A quantum enhancement in precision is of great importance in metrology such as
for imaging and microscopy. Recently, the quantum enhanced imaging making use of point
estimation theory is presented based on single parameter estimation procedure through the Fisher information approach \cite{PhysRevLett.109.123601}.
Since phase imaging is inherently a multiple parameter estimation problem,
the multiple phase estimation is of interest \cite{Humphreys2013}.
It is found that for unitary evolutions, simultaneous estimation (SE)
of multiple phases provides an advantage scaling $O(d)$ in the variance of the
estimation over individual estimation (IE) of each phase, where $d$ is number of phases to be estimated.
This conclusion holds for noiseless processes. However, in a realistic scenario, noise cannot be avoided due to
decoherence. An investigation of whether this advantage still exists for a general evolution is necessary.

For noisy processes, it is not known in general if and when the quantum enhancement of precision from $1/\sqrt{N}$ to $1/N$
can be achieved though general expressions for the uncertainty in the estimation are known.
The problem is that their calculation involves complex optimization procedures.
Fortunately, a general framework is proposed recently to obtain attainable and useful lower bound of the quantum Fisher information (QFI)
in noisy systems \cite{escher2011general}. In particular, this lower bound captures the main features
of the transition from the $1/N$ to $1/\sqrt {N}$ precisions for the cases of noisy channels
such as photon loss and dephasing. Those results are for the single parameter estimation.

In this work, we present a general framework for the {\it estimation of multiple phases with noise}.
We apply this framework to study a specific example of the photon loss type noise.
Photon loss is a very usual noise type in optical systems.
We make a conjecture that with only photon loss considered
the QFI matrix of the phases can be saturated for a certain set of initial probes, which means that we are in principal able to find a measurement $M$ to make the
Fisher information matrix after measurement equal to the QFI matrix. In this way, the QFI bound computed is a tight lower bound of the uncertainty of the estimation.
We show that in the limit of noiseless, the precision can achieve the Heisenberg limit $1/N$ with an
advantage of $O(d)$ for multiple phase, thus recover the known results \cite{Humphreys2013}.
With noise increasing, SE is always better than IE, but the $O(d)$ advantage may disappear asymptotically, with photon loss taken as an example.
At the same time, the precision of estimation decreases to the standard quantum limit (SQL) $1/\sqrt {N}$.
So similar as for single phase, our result of multiphase can also capture the main features of the
transition from Heisenberg limit to standard quantum limit.

We shall consider a multiple phase estimation model described by Fig.1. In the preparation stage, a probe state is created of the form
\begin{equation}
|\psi_0\rangle = \sum_{k=1}^{D} \alpha_{k}|N_{k, 0}, N_{k, 1}, \cdots, N_{k, d}\rangle = \sum_{k=1}^{D} \alpha_{k}|\bm{N}_k \rangle.
\label{initialState}
\end{equation}
We assume that the amount of resources employed in the estimation process is restricted by the photon number $N$, and $\bm{N}_k$ describes the $k$th possible distribution of $N$ photons in different modes, which is represented by a vector $(N_{k, 0}, N_{k, 1}, \cdots, N_{k, d})$ , where $N_{k, i}$ stands for the number of photons employed in the $i$th mode and $\sum_{i=0}^d N_{k,i}=N$. $D=(N+d)!/N!d!$ stands for the total number of possible distributions. Normalization is required such that $\sum_{k=1}^D|\alpha_k|^2=1$. In an estimation scheme, the probe state is chosen beforehand, and one aim of metrology is to find out the optimal probe to estimate the parameters. For simplicity we only choose pure states as probes, so we have $\rho_0 = |\psi_0\rangle\langle \psi_0|$.

In the evolution stage, we consider the case that states in different modes evolve independently. In the mode $i$, evolution is determined by the parameter $\theta_{i}$, expressed in terms of Kraus operators $\hat{\Pi}_{l_{i}}^{(i)}(\theta_{i})$, which satisfies $\sum_{l_{i}}\hat{\Pi}_{l_{i}}^{(i)\dagger}(\theta_{i})\hat{\Pi}_{l_{i}}^{(i)}(\theta_{i}) = \mathbb{I}$. The evolved state is then given by
\begin{equation}
\rho(\bm{\theta}) = \sum_{\bm{l}} \hat{\Pi}_{\bm{l}}(\bm{\theta}) \rho_0 \hat{\Pi}_{\bm{l}}^{\dagger}(\bm{\theta}),
\end{equation}
where we denote $\bm{\theta} = (\theta_1, ..., \theta_d)$, $\bm{l} = (l_0, l_1, ..., l_d)$ and $\hat{\Pi}_{\bm{l}}(\bm{\theta}) = \hat{\Pi}_{l_0}^{(0)}\otimes\hat{\Pi}_{l_1}^{(1)}(\theta_1)\otimes \cdots \otimes \hat{\Pi}_{l_d}^{(d)}(\theta_d).$

\section*{Results}

\subsection{The advantage of simultaneous estimation.}

As is shown \cite{Humphreys2013}, SE provides an $O(d)$ advantage over IE, without noise considered.
Here we shall show that even under general evolution, SE is still better than IE, but the $O(d)$ advantage may disappear gradually,
with photon loss taken as an example. We remark that our results of noisy processes can recover
the case of noiseless in a continuum manner thus possess the SE advantage.

In Fig. 1, only one reference mode $0$ is implemented to estimate the $d$ phases $\theta_1$ to $\theta_d$. We now consider the scheme to implement $d$ reference modes, with each connected to a corresponding phase. The initial state can be written as
\begin{equation}
|\psi_0\rangle = \sum_{k} \alpha_{k}|N_{k, 01},\cdots, N_{k, 0d}, N_{k, 1}, \cdots, N_{k, d}\rangle,
\end{equation}
where each reference mode experiences the same evolution as the original mode $0$. We remark that any IE strategy is equivalent to use an initial state with the form
\begin{equation}
|\psi_0\rangle = |\psi_1\rangle_{01, 1} \otimes |\psi_2\rangle_{02, 2} \otimes \cdots \otimes |\psi_d\rangle_{0d, d},
\end{equation}
and only separate measurement for each phase is allowed. Now we see that IE is actually contained in the complete set of SE strategies, which leads to the conclusion that SE is generally better than IE even under noise.

\subsection{Phase estimation under photon loss.}
A beam splitter is generally used to model photon loss. A possible set of Kraus operators in each mode is given by \cite{escher2011quantum}
\begin{equation}
\hat{\Pi}_{l_i}^{(i)} = \sqrt{\frac{(1-\eta_i)^{l_i}}{l_i !}}e^{i\theta_i \hat{n}_i} \eta_i^{\frac{\hat{n}_i}{2}}\hat{a_i}^{l_i},
\end{equation}
where $\eta$ is the square of the transmissivity $r$ (ranging from $\eta = 1$, lossless case, to $\eta = 0$, complete loss). It is conjectured in Supplementary Material that, as long as all the $\alpha_k$ in Eq.(\ref{initialState}) are real, for this particular set of Kraus operators, the QFI bound can be saturated.  Since equivalent sets of Kraus operators lead to the same evolved state, the QFI matrix should be the same no matter what Kraus operators are chosen. Consider the following set of Kraus operators
\begin{equation}
\hat{\Pi}_{l_i}^{(i)} = \sqrt{\frac{(1-\eta_i)^{l_i}}{l_i !}}e^{i\theta_i (\hat{n}_i - \delta_i l_i)} \eta_i^{\frac{\hat{n}_i}{2}}\hat{a_i}^{l_i},
\end{equation}
where $\delta_i$ are arbitrary real numbers that we are free to choose.

In the methods part, we have derived a method to give a lower bound for the optimal precision of multiple phase estimation
\begin{equation}
Cov(\bm{\theta}) \geq \frac{1}{C_Q(\bm{\theta}, \hat{\Pi}_{\bm{l}})},
\end{equation}
where the element of the matrix of $C_Q$ is
\begin{equation}
C_Q(\bm{\theta}, \hat{\Pi}_{\bm{l}})_{ij} = 4\{ \langle \hat{B}^{(ij)} \rangle_0 - \langle \hat{A}^{(i)}\rangle_0 \langle \hat{A}^{(j)}\rangle_0 \},
\end{equation}
with $\langle\cdots\rangle_0$ standing for ${}_S\langle \psi_0|\cdots|\psi_0 \rangle_S$ and
\begin{equation}
\hat{A}^{(i)} = \sum_{l_{i}}i\frac{d \hat{\Pi}_{l_{i}}^{(i)\dagger}}{d \theta_{i}}\hat{\Pi}_{l_{i}}^{(i)},
\end{equation}
\begin{equation}
\hat{B}^{(ij)} =
\begin{cases}
 \sum_{l_{i}} \frac{d \hat{\Pi}_{l_{i}}^{(i)\dagger}}{d \theta_{i}} \frac{d \hat{\Pi}_{l_{i}}^{(i)}}{d \theta_{i}},  i = j \\
\hat{A}^{(i)}\hat{A}^{(j)},  i \neq j
\end{cases}
\end{equation}

Under the noise of photon loss, following the same calculation as in the single phase case, we have \cite{escher2011general}
\begin{eqnarray}
&& \hat{A}^{(i)} = a_i\hat{n}_i, \nonumber \\
&& \hat{B}^{(ij)} =
\begin{cases}
\hat{A}^{(i)}\hat{A}^{(j)}, i \neq j \\
a_i^2\hat{n}_i^2+b_i\hat{n}_i, i = j ,
\end{cases}
\label{AiBi}
\end{eqnarray}
with $a_i = 1-(1+\delta_i)(1-\eta_i)$, $b_i = (1+\delta_i)^2\eta_i(1-\eta_i)$.
For simplicity of calculation, we suppose that $\eta_i = \eta$ for all $i$, or all modes are symmetric.

We first consider the best IE strategy to estimate $d$ phases with limited resources of $N$ photons. Generally the minimum uncertainty of the estimate of phase $i$ can be written as
\begin{equation}
\Delta \theta_i^2 = \frac{C_t}{n_i^t},
\label{IEminuncer}
\end{equation}
where $t$ is the scaling coefficient under certain conditions with $t=2$ being the Heisenberg scale and $t=1$ being the SQL scale. $C_t$ is a constant and $n_i$ is the number of photons employed in the estimation of phase $i$. Since all modes are symmetric, we assume that under the best IE strategy, the uncertainty of each phase follows the same scaling. We then need to minimize $\sum_{i=1}^d \Delta \theta_i^2 = C_t \sum_{i=1}^d \frac{1}{n_i^t}$. Through basic calculation we know that the minimum is obtained when the estimation of each phase uses the same amount of resources, which is $N/d$ photons, for any positive $t$. Then we have
\begin{equation}
\min \sum_{i=1}^d \Delta \theta_i^2 = \frac{C_t}{(N/d)^t}d.
\label{minIE}
\end{equation}

Now we turn to the SE strategy.
If we choose $\delta_i = \frac{1}{1-\eta_i} - 1$ and substitute them into Eq.(\ref{AiBi}), all the off-diagonal terms of $C_Q$ will disappear, we then have
\begin{equation}
Tr[C_Q^{-1}] = \sum_i \frac{4\eta}{1-\eta} \frac{1}{\langle \hat{n}_i \rangle_0},
\label{CQ-1}
\end{equation}
from which we can clearly observe the disappearance of the Heisenberg scale as is expected.

To see that the $O(d)$ advantage may disappear in the asymptotic case, we first assume that $N \gg 1$, $d \gg 1$, $N/d \gg 1$. From Eq.(\ref{CQ-1}), we are to seek a state $|\psi_0\rangle$ which maximizes $\sum_i \frac{1}{\langle \hat{n}_i \rangle_0}$. Since $\sum_i \langle \hat{n}_i \rangle_0 \leq N$, we have $\sum_i \frac{1}{\langle \hat{n}_i \rangle_0} \leq  \frac{d^2}{N}$ and the equality is attained when $\langle \hat{n}_i \rangle_0 = \frac{N}{d}$ for any $i$. Then a lower bound for SE is obtained:
\begin{equation}
Tr[Cov(\bm{\theta})] \geq \frac{1-\eta}{4 \eta}\frac{d^2}{N}.
\label{SE}
\end{equation}

We know in the asymptotic case, the scaling coefficient $t$ in Eq.(\ref{minIE}) is 1, and the total variance is
\begin{equation}
\sum_i \Delta \theta_i^2 \approx  C_1\frac{1}{N/d} d = C_1\frac{d^2}{N}.
\label{IE}
\end{equation}
Compare Eq.(\ref{SE}) and Eq.(\ref{IE}), we see that the $O(d)$ advantage no longer exists.

In order to exhibit more clearly the transition from the Heisenberg scale with the $O(d)$ advantage to the SQL scale without the $O(d)$ advantage, we investigate the SE strategy using a specific probe state $|\psi_s\rangle$. $|\psi_s\rangle$ is a generalized $N00N$ state as defined in Ref.\cite{Humphreys2013}, explicitly written as
\begin{eqnarray}
|\psi_s\rangle = && \alpha(|0,N,0,\cdots,0\rangle + |0,0,N,\cdots,0\rangle+\cdots+ \nonumber \\ && |0,0,0,\cdots,N\rangle)+\beta|N,0,0,\cdots,0\rangle,
\end{eqnarray}
where $\alpha^2=\frac{1}{d+\sqrt{d}}$ and $d\alpha^2+\beta^2 = 1$. The reason we choose this state is that in the noiseless case estimation with this state has both the Heisenberg scale and the O(d) advantage\cite{Humphreys2013}, and we will show how they disappear as noise becomes significant. To further simplify the calculation, we assume that $\delta_i = \delta$, which is reasonable since all modes are symmetric. Then only one variable $\delta$ needs to be optimized to make the lower bound $C_Q$ as tight as possible. Asymptotically we have
\begin{equation}
Tr[C_Q^{-1}] \approx \frac{1}{4} \frac{1}{ \frac{1}{(\frac{1-\eta}{\eta} N+1)^2}(\frac{N}{d})^2 + \frac{\frac{1-\eta}{\eta} N^2}{(\frac{1-\eta}{\eta} N+1)^2}\frac{N}{d}\frac{1}{d}},
\end{equation}
when $\delta = \frac{N/\eta}{\frac{1-\eta}{\eta}N+1}-1$, see Supplementary Material for details. For $\frac{1-\eta}{\eta}\ll \frac{1}{N}$,
we have
$$Tr[C_Q^{-1}] = \frac{1}{4} \frac{1}{(N/d)^2}.$$
We see that it is the Heisenberg scale, additionally,
compared with Eq.(\ref{minIE}), the $O(d)$ advantage of SE exists. Whereas for $N\gg\frac{\eta}{1-\eta}$, we have $Tr[C_Q^{-1}] = \frac{1-\eta}{4\eta} \frac{1}{N/d} d$. We see that it is the SQL scale and compared with Eq.(\ref{minIE}), the $O(d)$ advantage of SE disappears.

Although we have proven that SE provides at most a constant increase of precision over IE
asymptotically for large noise, it doesn't mean that there is no need to use the SE strategy. Rather contrarily, it is shown in Fig.[2] that for $d = 2$, $\eta = 0.9$ and small numbers $N$, a significant decrease of uncertainty about $50\%$ can be achieved. For IE, an optimized state over all states of the form $\sum_{n = 0}^{N/d} \alpha_n |n, N/d-n\rangle$ is chosen as the probe to estimate an individual phase. We have calculated  a lower bound of the QFI \cite{escher2011general}. For SE without loss, the state $|\psi_s\rangle$ is chosen as the probe. For SE with loss, we use the same probe and calculate its QFI matrix numerically. Since we have proven that for this initial state, the QFI matrix can be locally saturated, we have $|\Delta \bm{\theta}_{SE}|^2 = Tr[I_Q(\bm{\theta})^{-1}]$. So in principal, an advantage of SE over IE larger than that shown in Fig.[2] can be obtained. From the result, we see that if we need to estimate multiple phases,
we should estimate simultaneously to achieve higher precision.

In Fig. 2, we have also made a comparison of different estimation strategies versus various $\eta$. We see that under low $\eta$, which means the photon loss is significant, SE using states $|\psi_s\rangle$ is worse than IE. This is understandable, because for calculating $|\Delta \bm{\theta}_{IE}|^2$, we have used an optimal probe, but for calculating $|\Delta \bm{\theta}_{SE}|^2$, only $|\psi_s\rangle$ is used. $|\psi_s\rangle$ is a generalized $N00N$ state and is vulnerable to photon loss. A state robust against photon loss may be necessary \cite{ZhangYLPRA2}.
For higher $\eta$, $|\psi_s\rangle$ is enough to beat the IE strategy.

\section*{Discussion}

We have presented a lower bound for the error in multi-parameter estimation under noise, within the framework of quantum metrology, and photon loss is exemplified. We have proved the usefulness of this bound by showing that it can capture the main feature of the transition from the Heisenberg limit with the $O(d)$ advantage to the SQL limit without the $O(d)$ advantage as noise becomes significant. We have also shown the advantage of SE over IE in precision.
The enhancement in precision can also be applied for single phase
by replicating it to several copies.
This novel scheme is better than simply duplicating the measurement instrument.
Our analysis of multiple phase estimation should be of wide interest in many problems. Quantum enhanced phase imaging is one potential application. A recent investigation of quantum phase imaging used point estimation with single parameter\cite{PhysRevLett.109.123601}, since phase imaging is inherently a multiparameter estimation problem, our results provide an approach to this problem.
Our results should also be of interest in gravitational wave detection\cite{LIGO}, since it can be recast as optical phase estimation\cite{Giovannetti2011}.
They will also motivate an investigation into the role of noise in quantum enhancement.
Thus, the application of our results is worth investigating for various
quantum metrology problems.

\begin{methods}

It is known that, the precision of the estimate of $\bm{\theta}$, described by its covariance matrix $Cov(\bm{\theta})$, is limited by the quantum Cram\'{e}r-Rao (QCR) inequality \cite{Helstrom1976, Paris2009}
\begin{equation}
Cov(\bm{\theta}) \geq (M I_Q(\bm{\theta}))^{-1},
\end{equation}
where the inequality means that $Cov(\bm{\theta}) - (M I_Q(\bm{\theta}))^{-1}$ is positive semidefinite, $I_Q(\bm{\theta})$ is the QFI matrix, $M$ is the repetition of the whole estimation process. Here we have assumed that the estimator of $\bm{\theta}$ is unbiased. This is a reasonable assumption since Cram\'{e}r has proved that the maximum likelihood method will give an asymptotic unbiased estimate as $M\rightarrow \infty$ \cite{cramer1946}.
A brief introduction about the QFI approach for quantum metrology is presented in Supplementary Material. Since we are interested only in the quantum enhancement, we shall set $M$ to $1$ for this letter.
The total variance of all the phases is then
\begin{equation}
|\Delta \bm{\theta}|^2 = \sum_{i=1}^d \delta \theta_i^2 = Tr[Cov(\bm{\theta})] \geq Tr[I(\bm{\theta})^{-1}].
\end{equation}

Inspired by the work \cite{escher2011general}, we propose a general method to derive an upper bound $C_Q(\bm{\theta}, \hat{\Pi}_{\bm{l}})$ of $I_Q(\bm{\theta})$, where $\hat{\Pi}_{\bm{l}}$ is any Kraus representation of the quantum channel.
Suppose the real value of the parameter vector is $\bm{\theta}$, and $\bm{\epsilon}$ is an infinitesimal increment, then we have the relation between the Bures fidelity and the QFI matrix at $\bm{\theta}$ \cite{Paris2009}:
\begin{equation}
(F_B[\rho(\bm{\theta}), \rho(\bm{\theta}+\bm{\epsilon})])^2 = 1 - \frac{1}{4} \sum_{i, j} \epsilon_i\epsilon_j I_Q(\bm{\theta})_{ij},
\end{equation}
where the Bures fidelity is defined as: $F_B[\rho, \sigma] = Tr\sqrt{\sqrt{\rho}\sigma\sqrt{\rho}}$.
Uhlmann's theorem states that \cite{Nielsen2001quantum}
\begin{equation}
(F_B[\rho(\bm{\theta}), \rho(\bm{\theta}+\bm{\epsilon})])^2 = \max_{|\Psi(\bm{\theta}+\bm{\epsilon})\rangle}|\langle \Phi(\bm{\theta})|\Psi(\bm{\theta}+\bm{\epsilon})\rangle|^2,
\end{equation}
where $|\Phi(\bm{\theta})\rangle$ is an arbitrary purification of $\rho(\bm{\theta})$ in an enlarged space $SE$, and $|\Psi(\bm{\theta}+\bm{\epsilon})\rangle$ runs over all purifications of $\rho(\bm{\theta}+\bm{\epsilon})$. Since $|\langle \Phi(\bm{\theta})|\Psi(\bm{\theta}+\bm{\epsilon})\rangle|^2 = F_B(|\Phi(\bm{\theta})\rangle, |\Psi(\bm{\theta}+\bm{\epsilon})\rangle)^2 = 1-\frac{1}{4}\sum_{i, j}\epsilon_i\epsilon_j C_Q(\bm{\theta})_{ij}$, where $C_Q(\bm{\theta})$ is the QFI matrix at $\bm{\theta}$ in space $SE$, we have $I_Q(\bm{\theta})\leq C_Q(\bm{\theta})$. The equality may actually be achieved. Because for pure states $|\Psi(\bm{\theta})\rangle$, its QFI matrix can be explicitly written out. This will provide us a method to derive useful analytical bounds of $I_Q(\bm{\theta})$.

Notice that for the scheme of Fig. 1, although the probe state may be correlated, the evolution is separated for different modes. Thus rather than to purify the system $S$ on the whole, we may purify each mode independently, which greatly reduces the difficulty of purification. Add an environment $E_i$ to the respect system $S_i$, and purify the evolution $\left\{ \hat{\Pi}_{l_i}^{(i)}\right\}$ to a unitary one $\hat{U}_i^{(S_i E_i)}$, the evolved state $\rho_S(\bm{\theta})$ becomes a pure state $\left|\Psi(\bm{\theta})\right\rangle_{SE}$, given by
\begin{equation}
|\Psi({\bm{\theta})}\rangle_{SE} = \hat{U}^{(SE)}(\bm{\theta})|\psi_0\rangle_S|0\rangle_E,
\end{equation}
where $\hat{U}^{(SE)}(\bm{\theta}) = \hat{U}_0^{(S_0E_0)}\otimes _{i=1}^d
\hat{U}_i^{(S_i E_i)}(\theta_i)$,
$|0\rangle_E = \otimes_{i=0}^d|0\rangle_{E_i}$.
The purified unitary evolution is connected to the original Kraus representation through the equation \cite{Nielsen2001quantum},
\begin{equation}
\hat{\Pi}_{l_i}^{(i)}(\theta_i) = {}_{E_i}\langle l_i| \hat{U}^{(S_iE_i)}_{i}(\theta_i) | 0\rangle_{E_i},
\end{equation}
where $|l_i\rangle_{E_i}$ form a basis for the environment $E_i$.

We show in Supplemental Material  that the QFI matrix for the enlarged total system SE can then be expressed as
\begin{equation}
C_Q(\bm{\theta}, \hat{\Pi}_{\bm{l}})_{ij} = 4\{ \langle \hat{B}^{(ij)} \rangle_0 - \langle \hat{A}^{(i)}\rangle_0 \langle \hat{A}^{(j)}\rangle_0 \},
\label{CQ}
\end{equation}
with $\langle\cdots\rangle_0$ standing for ${}_S\langle \psi_0|\cdots|\psi_0 \rangle_S$ and
\begin{equation}
\hat{A}^{(i)} = \sum_{l_{i}}i\frac{d \hat{\Pi}_{l_{i}}^{(i)\dagger}}{d \theta_{i}}\hat{\Pi}_{l_{i}}^{(i)},
\end{equation}

\begin{equation}
\hat{B}^{(ij)} =
\begin{cases}
 \sum_{l_{i}} \frac{d \hat{\Pi}_{l_{i}}^{(i)\dagger}}{d \theta_{i}} \frac{d \hat{\Pi}_{l_{i}}^{(i)}}{d \theta_{i}},  i = j \\
\hat{A}^{(i)}\hat{A}^{(j)},  i \neq j
\end{cases}
\end{equation}

So at first place, we have $I_Q(\bm{\theta}) = \min_{\hat{\Pi}_{\bm{l}}}  C_Q(\bm{\theta}, \hat{\Pi}_{\bm{l}})$, with the minimization running over all possible Kraus representations of the quantum channel. In order to reduce the difficulty of the optimization process, we only consider independent purification of each mode, such that $\hat{\Pi}_{\bm{l}}(\bm{\theta}) = \hat{\Pi}_{l_0}\otimes\hat{\Pi}_{l_1}(\theta_1)\otimes\cdots\otimes\hat{\Pi}_{l_d}(\theta_d)$. Further we can restrict the minimization process to a subclass of all the possible $\hat{\Pi}_{\bm{l}}$, depending on a few variational parameters which shall be optimized. The subclass may be constructed based on physical insight. In this way nontrivial bound can also be obtained as we will present below.

\end{methods}

\begin{addendum}

\item [Acknowledgement]

This work is supported by ``973" program (2010CB922904), NSFC (11175248), and
grants from Chinese Academy of Sciences.

\item [Author Contributions]
J.-D. Y. and H. F. proposed the model. J.-D. Y.
calculates the results. J.-D. Y. and Y.-R. Z. analyzed the results. J.-D. Y. and H.F. wrote the paper.

\item [Competing Interests]
The authors declare that they have no competing financial interests.

\item [Correspondence]
Correspondence and requests for materials should be addressed to
H.F. or J.-D. Y.
\end{addendum}

\newpage

\textbf{Figure 1. A multiple phase estimation model.} An initially prepared probe state $|\psi_0\rangle$ undergoes a general evolution described by $d+1$ sets of Kraus operators, depending on $d$ parameters which we are supposed to estimate simultaneously. Different modes evolve independently.

\textbf{Figure 2. A comparison of SE and IE strategies for multiple phase estimation with $d = 2$, $\theta_1=2$, $\theta_2 = 2$.  }
For (a), $\eta$ is fixed at $0.9$ and $N$ is various. For (b), $N$ is fixed at $6$ and $\eta$ is various. The black solid line gives the total variance $|\Delta \bm{\theta}_{SEideal}|^2$ without any noise using the probe states $|\psi_s\rangle$. The red dashed line gives the total variance $|\Delta \bm{\theta}_{SE}|^2$ under photon loss using the probe states $|\psi_s\rangle$. The blue dotted line gives a lower bound of the total variance $|\Delta \bm{\theta}_{IE}|^2$ under photon loss using IE strategy with the optimal probe .

\newpage
\begin{figure}[tbp]
\begin{center}
\epsfig{file=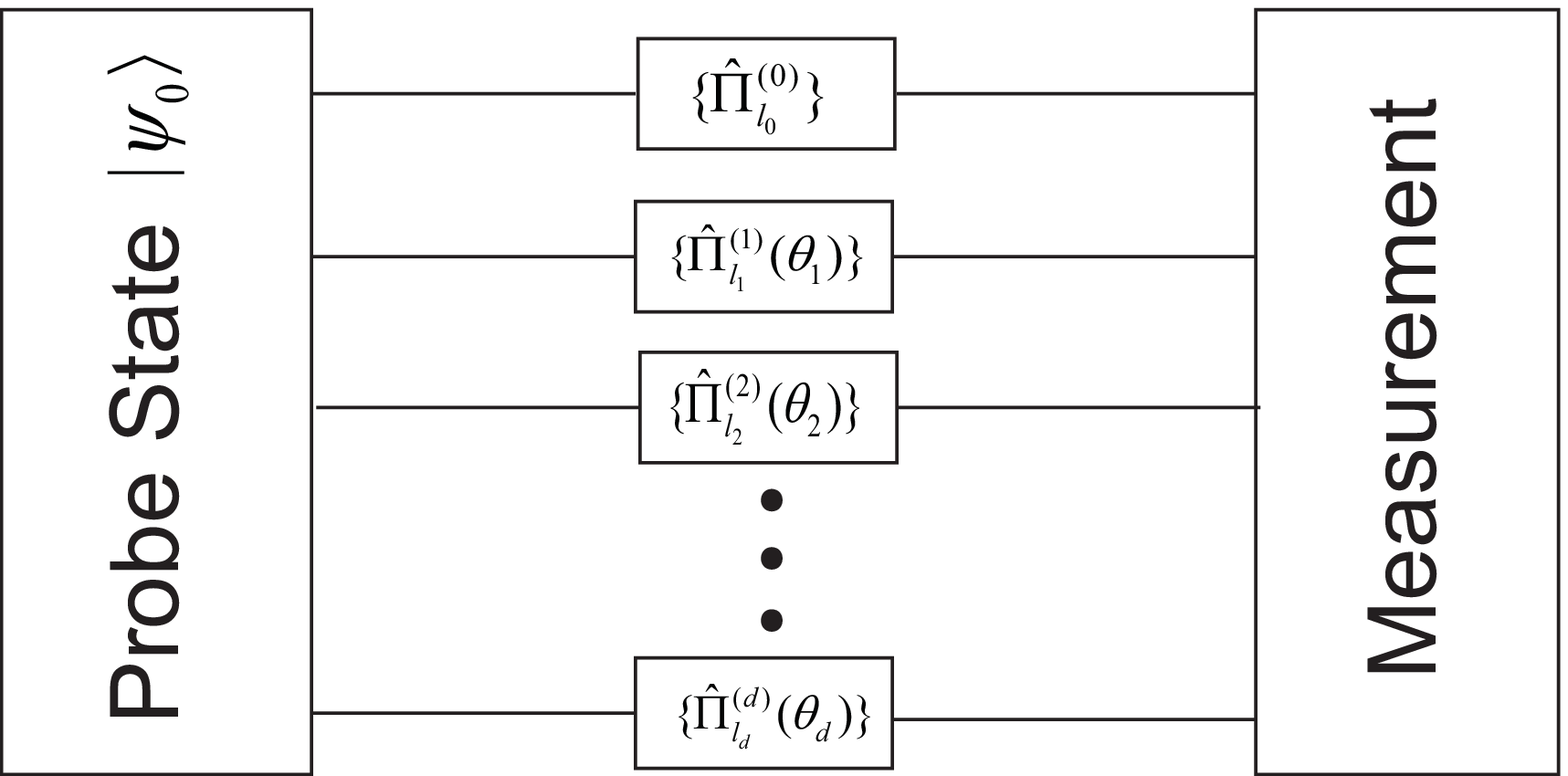,width=16cm}
\end{center}
\par
\label{fig1}
\end{figure}

\begin{center}
Figure 1
\end{center}

\newpage
\begin{figure}[tbp]
\begin{center}
\epsfig{file=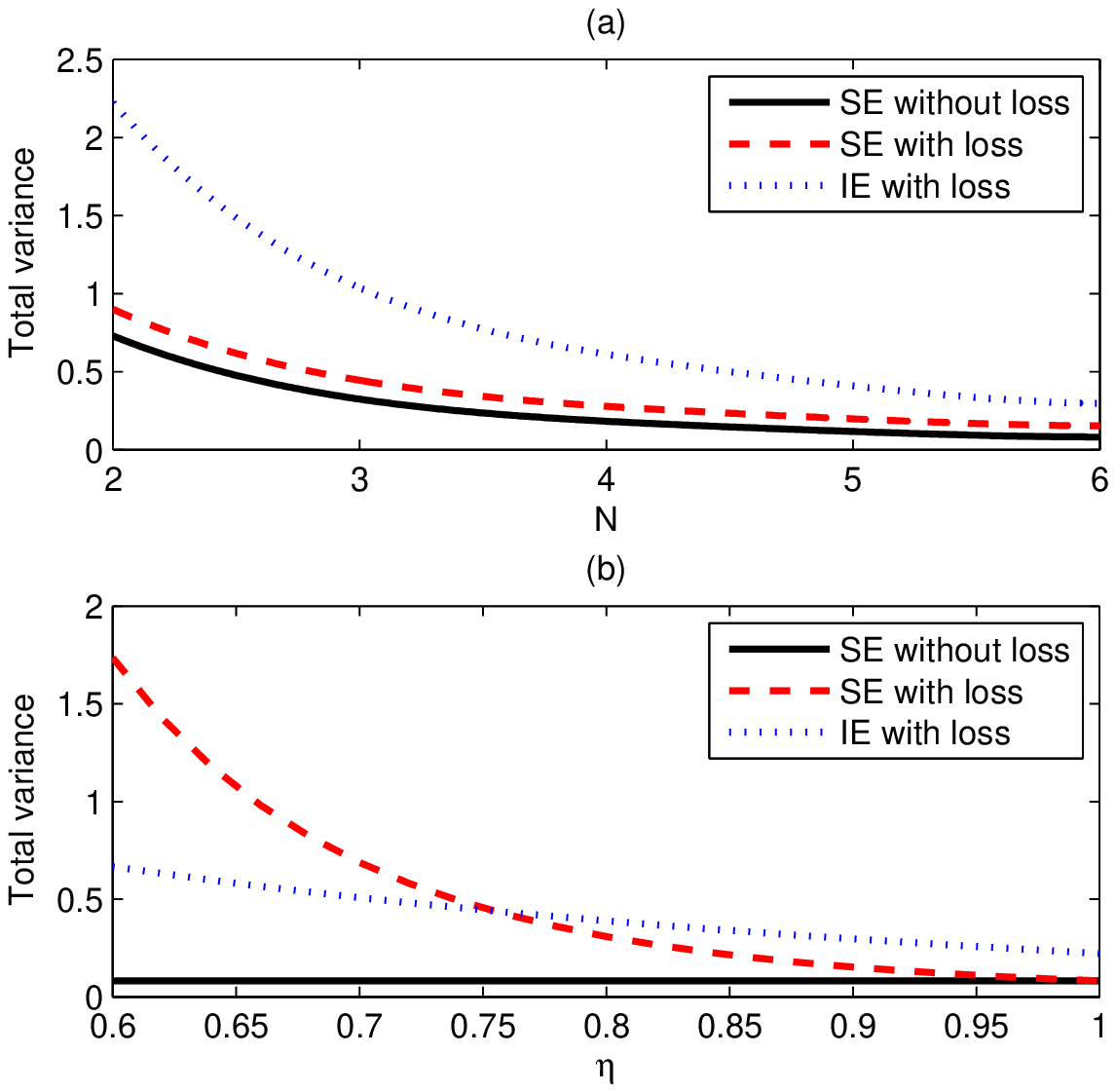,width=16cm}
\end{center}
\par
\label{fig2}
\end{figure}

\begin{center}
Figure 2
\end{center}

\end{document}